\documentclass[a4paper]{jpconf}
\usepackage{graphicx}
\usepackage{amsmath,amsfonts,amssymb}
\usepackage{euscript}
\usepackage{array}
\usepackage{longtable}
\renewcommand{\d}{\mathrm{d}}

\begin{document}
\title{Light-like $\kappa$-deformations and scalar field theory via Drinfeld twist}

\author{Tajron Juri\'c$^1$, Stjepan Meljanac$^2$ and Andjelo Samsarov$^3$ }

\address{$^{1,2}$ Rudjer Bo\v skovi\'c Institute, Theoretical Physics Division, Bijeni\v cka  c.54, HR-10002 Zagreb,
Croatia}
\address{$^3$ Dipartimento di Matematica e Informatica, Universita di Cagliari,
viale Merello 92, 09123 Cagliari, Italy and INFN, Sezione di Cagliari}

\ead{$^1$ tjuric@irb.hr, $^2$ meljanac@irb.hr, $^3$ samsarov@unica.it}

\begin{abstract}
In this article we will use the Drinfeld twist leading to light-like  $\kappa$-deformations of Poincar\'e algebra. We shall apply the standard Hopf algebra methods in order to define the star-product, which shall be used  to formulate a scalar field theory compatible with $\kappa$-Poincar\'e-Hopf algebra. Using this approach we show that there is no problem with formulating integration on $\kappa$-Minkowski space and no need for introducing a new measure. We have shown that the $\star$-product obtained from this twist  enables us to define a free scalar field theory on $\kappa$-Minkowski space that is equivalent to a commutative one on a usual Minkowski space. We also discuss the interacting $\phi^4$ scalar field model compatible with $\kappa$-Poincar\'e-Hopf algebra.
\end{abstract}

\section{Introduction}
Quantum field theory is a framework that describes the multi-particle systems in accordance with the quantum mechanical principles and the principles of special relativity. Therefore, the Poincar\'e symmetry is in the very heart of quantum field theory. It is argued that at very high energies the gravitational effects can no longer be neglected and that the spacetime is no longer a smooth manifold, but rather a fuzzy, or better to say a noncommutative space \cite{doplicher}. The quantum field theory on such noncommutative manifolds requires a new framework. This new framework is provided by noncommutative geometry \cite{connes}. In this framework, the search for diffeomorphisms that leave the physical action invariant leads to deformation of Poincar\'e symmetry, with $\kappa$-Poincar\'e symmetry being among the most extensively studied \cite{Lukierski, Majid, Kowalski}.

$\kappa$-deformed Poincar\'e symmetry is algebraically described by the $\kappa$-Poincar\'e-Hopf algebra and is an example of deformed relativistic symmetry that can possibly describe the physical reality at the Planck scale. $\kappa$ is the deformation parameter usually interpreted as the Planck mass or some quantum gravity scale.

The deformation of the symmetry group is  realized through the application of the Drinfeld twist on that group \cite{chaichan}. In order to implement this twisted (deformed) Poincar\'e symmetry into the quantum field theory,  it is necessary to implement
twisted statistics \cite{Balachandran, Govindarajan}, with the form of the interaction also being dictated by the quantum symmetry. The virtue of the twist
formulation is that the deformed (twisted) symmetry algebra is the same as the original undeformed one and only the coalgebra
structure changes, leading to the same free field structure as in the  corresponding commutative field theory.

There have been attempts in the literature to obtain $\kappa$-Poincar\'e-Hopf algebra from the Abelian \cite{Meljanac, Govindarajan} and Jordanian twist \cite{Borowiec11}, but the problem with these twists is that they can not be expressed in terms of the Poincar\'e generators only. The $\kappa$-Poincar\'e-Hopf algebra was obtained using a twist in a Hopf algebroid approach in \cite{Juric2}. Particularly, a full description of deformation of Poincar\'e algebra in terms of  the
twist and the R-matrix is missing. However, for the light-like  $\kappa$-deformations of Poincar\'e algebra there is a Drinfeld twist \cite{Juric2, Juric1, Juric3}. It was considered earlier in the literature \cite{Kulish, Borowiec}, but it was written only in the light-cone basis.

In this paper we will use the Drinfeld twist leading to light-like $\kappa$-deformations of Poincar\'e algebra \cite{Juric2, Juric1, Juric3} and apply the methods developed in \cite{Drinfeld, Drinfeld1, Aschieri, Aschieri1, Aschieri2} in order to first define the star-product and then to use it with the purpose of formulating a scalar field theory compatible with $\kappa$-Poincar\'e-Hopf algebra. As it turns out that the $\kappa$-Poincar\'e algebra acts on the $\kappa$-Minkowski space in the analogous way as the Poincar\'e algebra acts on the usual Minkowski space, it appears that there is no problem with formulating integration on $\kappa$-Minkowski space and no need for introducing a new measure like in \cite{Felder, Dimitrijevic2, Dimitrijevic, Moller, Agostini}. The star product stemming from our Drinfeld twist has the cyclic property under the sign of integration.

The paper is organized as follows. In section 2 we outline the basic definitions of Hopf algebra, Drinfeld twist and star product. In section 3 we apply the Hopf algebra methods to Poincar\'e algebra and by using our light-like Drinfeld twist \cite{Juric2, Juric1, Juric3}, derive the full  $\kappa$-Poincar\'e-Hopf algebra and the corresponding $\kappa$-Minkowski space. After proving some useful properties of the star-product, we formulate the free scalar field theory (in section 4) and a model of interacting $\phi^4$ field theory on $\kappa$-Minkowski space (in section 5). Finally, in section 6 we give our conclusion and future prospects.

\section{Hopf algebra and Drinfeld twist}
Let us first state some general facts about Drinfeld twists and Hopf algebras. Consider a Lie algebra $\mathfrak{g}$ and its universal enveloping algebra $\mathcal{U}(\mathfrak{g})$. $\mathcal{U}(\mathfrak{g})$ is an associative unital algebra. It is also a Hopf algebra with coproduct $\Delta:\mathcal{U}(\mathfrak{g})\longrightarrow\mathcal{U}(\mathfrak{g})\otimes\mathcal{U}(\mathfrak{g})$, counit $\epsilon:\mathcal{U}(\mathfrak{g})\longrightarrow\mathbb{C}$ and antipode $S:\mathcal{U}(\mathfrak{g})\longrightarrow\mathcal{U}(\mathfrak{g})$ satisfying 
\begin{equation}\begin{split}
(\Delta\otimes 1)\Delta(\xi)&=(1\otimes\Delta)\Delta(\xi),\\
m\left[(\epsilon\otimes 1)\Delta(\xi)\right]&=m\left[(1\otimes\epsilon)\Delta(\xi)\right]=\xi,\\
m\left[(S\otimes 1)\Delta(\xi)\right]&=m\left[(1\otimes S)\Delta(\xi)\right]=\epsilon(\xi)1,
\end{split}\end{equation}
where $m:\mathcal{U}(\mathfrak{g})\otimes\mathcal{U}(\mathfrak{g})\longrightarrow\mathcal{U}(\mathfrak{g})$ is the multiplication map, $1$ is the unity in $\mathcal{U}(\mathfrak{g})$  and $\xi\in\mathcal{U}(\mathfrak{g})$. For the generator $g\in\mathfrak{g}$ we have
\begin{equation}\begin{split}
\Delta(g)=g\otimes 1 + 1\otimes g&, \quad \Delta(1)=1\otimes 1,\\
\epsilon(g)=0&, \quad \epsilon(1)=1,\\
S(g)=-g&, \quad S(1)=1,
\end{split}\end{equation}
and this extends to the whole $\mathcal{U}(\mathfrak{g})$ by requiring that $\Delta$ and $\epsilon$ are homomorphisms and linear, while $S$ is an antihomomorphism and also linear. 

The Drinfeld twist $\mathcal{F}\in\mathcal{U}(\mathfrak{g})\otimes\mathcal{U}(\mathfrak{g})$ is an invertible element\footnote{Here we actually extend the notion of enveloping algebra to some formal power series in $\frac{1}{\kappa}$ (we replace the field $\mathbb{C}$ with the ring $\mathbb{C}[[\frac{1}{\kappa}]]$) and we correspondingly consider the Hopf algebra $(\mathcal{U}(\mathfrak{g})[[\frac{1}{\kappa}]], \Delta, S, \epsilon, m)$, but for the sake of brevity we shall denote $\mathcal{U}(\mathfrak{g})[[\frac{1}{\kappa}]]$  by $\mathcal{U}(\mathfrak{g})$. In this context we have an additional requirement on the twist $$\mathcal{F}=1\otimes 1+\mathcal{O}(\frac{1}{\kappa}).$$ } of the Hopf algebra $\mathcal{U}(\mathfrak{g})$ that satisfies the cocycle condition
\begin{equation}\label{coc}
(\mathcal{F}\otimes 1)(\Delta\otimes 1)\mathcal{F}=(1\otimes \mathcal{F})(1\otimes \Delta)\mathcal{F}
\end{equation}
and the normalization condition
\begin{equation}\label{norm}
m(\epsilon\otimes 1)\mathcal{F}=1=m(1\otimes \epsilon)\mathcal{F}.
\end{equation}
The cocycle condition \eqref{coc} is responsible for the associativity of the $\star$-product (to be defined in the next subsection). The twist operator $\mathcal{F}$ generates a new Hopf algebra $\mathcal{U}^{\mathcal{F}}(\mathfrak{g})$ given by $(\mathcal{U}^{\mathcal{F}}(\mathfrak{g}), \Delta^{\mathcal{F}}, \epsilon^{\mathcal{F}}, S^{\mathcal{F}}, m^{\mathcal{F}})$, where $\mathcal{U}^{\mathcal{F}}(\mathfrak{g})=\mathcal{U}(\mathfrak{g})$ as algebras and $\epsilon^{\mathcal{F}}=\epsilon$, while the new coproduct $\Delta^{\mathcal{F}}$ and antipode $S^{\mathcal{F}}$ are given by
\begin{equation}\begin{split}
\Delta^{\mathcal{F}}&=\mathcal{F}\Delta \mathcal{F}^{-1},\\
S^{\mathcal{F}}&=\chi S \chi^{-1},
\end{split}\end{equation}
where $\chi^{-1}=m\left[(S\otimes 1)\mathcal{F}^{-1}\right]$.

\subsection{$\star$-product}
Let us consider an algebra $\mathcal{A}$ (over $\mathbb{C}[[\frac{1}{\kappa}]]$) and an action of the Lie algebra $\mathfrak{g}$ on $\mathcal{A}$, $(g,a)\mapsto g(a)\equiv g \triangleright a$ defined for all $g\in\mathfrak{g}$ and $a\in\mathcal{A}$. The action of $\mathfrak{g}$ on $\mathcal{A}$ induces an action\footnote{For example the element $g'g\in\mathcal{U}(\mathfrak{g})$ has the action $g'(g(a))=g'\triangleright(g\triangleright a)$. Also note that $\triangleright:\mathcal{U}(\mathfrak{g})\otimes\mathcal{A}\longrightarrow\mathcal{A}$.} of the universal enveloping algebra $\mathcal{U}(\mathfrak{g})$ on $\mathcal{A}$. This way the algebra $\mathcal{A}$ is a $\mathcal{U}(\mathfrak{g})$-module algebra, that is, the algebra structure of the $\mathcal{U}(\mathfrak{g})$-module $\mathcal{A}$ is compatible with the $\mathcal{U}(\mathfrak{g})$ action
\begin{equation}
\xi\triangleright(ab)=\mu\left[\Delta(\xi)(\triangleright\otimes\triangleright)(a\otimes b)\right], \quad \xi\triangleright 1=\epsilon(\xi)1,
\end{equation}
 for all $\xi\in\mathcal{U}(\mathfrak{g})$ and $a,b\in\mathcal{A}$, where $1$ is the unit in $\mathcal{A}$, and $\mu$ is the multiplication map in the algebra $\mathcal{A}$.

For a given twist $\mathcal{F}\in\mathcal{U}(\mathfrak{g})\otimes\mathcal{U}(\mathfrak{g})$ satisfying \eqref{coc} and \eqref{norm} we can construct a deformed algebra $\mathcal{A}_{\star}$. The new algebra $\mathcal{A}_\star$ has the same vector space structure as $\mathcal{A}$, but the multiplication is changed via
\begin{equation}\label{star}
a\star b=\mu\left[\mathcal{F}^{-1}(\triangleright\otimes\triangleright)(a\otimes b)\right].
\end{equation}
The new product $\star$ is called the \textsl{star}-product and it is easy to see that it is associative if $\mathcal{F}$ satisfies \eqref{coc}. One can show that the algebra $\mathcal{A}_{\star}$ is a module algebra with respect to a deformed Hopf algebra $(\mathcal{U}^{\mathcal{F}}(\mathfrak{g}), \Delta^{\mathcal{F}}, \epsilon^{\mathcal{F}}, S^{\mathcal{F}}, m^{\mathcal{F}})$.

\section{$\kappa$-Poincar\'e-Hopf algebra and $\kappa$-Minkowski space via Drinfeld twist} 

Now we will consider the Poincar\'e algebra $\mathfrak{p}$ as the Lie algebra $\mathfrak{g}$ and apply the formalism outlined in the previous section. The Poincar\'e algebra is a semidirect product of Lorentz algebra and translation algebra, and is generated by generators of boosts $M_{0i}$, rotations $M_{ij}$ and translations $P_{\mu}$ satisfying
\begin{equation}\begin{split}
[M_{\mu\nu},M_{\lambda\rho}]&=-i(\eta_{\nu\lambda}M_{\mu\rho}-\eta_{\mu\lambda}M_{\nu\rho}-\eta_{\nu\rho}M_{\mu\lambda}+\eta_{\mu\rho}M_{\nu\lambda}),\\
[M_{\mu\nu},P_{\lambda}]&=-i(\eta_{\nu\lambda}P_{\mu}-\eta_{\mu\lambda}P_{\nu}).
\end{split}\end{equation}
The universal enveloping algebra of Poincar\'e algebra $\mathcal{U}(\mathfrak{p})$ has a Hopf algebra structure. For the generators we have
\begin{equation}\begin{split}
\Delta M_{\mu\nu}=M_{\mu\nu}\otimes 1+1\otimes M_{\mu\nu}&, \quad \Delta P_{\mu}=P_{\mu}\otimes 1+1\otimes P_{\mu}, \\
\epsilon(M_{\mu\nu})=0&, \quad \epsilon(P_{\mu})=0,\\
S(M_{\mu\nu})=-M_{\mu\nu}&, \quad S(P_{\mu})=-P_{\mu}.
\end{split}\end{equation}
For the algebra $\mathcal{A}$ we choose the algebra of functions on Minkowski space. Thus, for any $f\in\mathcal{A}$ we have the following action $\triangleright:\mathcal{U}(\mathfrak{p})\otimes\mathcal{A}\longrightarrow\mathcal{A}$ 
\begin{equation}\begin{split}
M_{\mu\nu}\triangleright f&=-i\left(x_{\mu}\frac{\partial f}{\partial x^{\nu}}-x_{\nu}\frac{\partial f}{\partial x^{\mu}}\right),\\
P_{\mu}\triangleright f&=-i\frac{\partial f}{\partial x^{\mu}}.
\end{split}\end{equation}

The Drinfeld twist $\mathcal{F}$ for the light-like $\kappa$-deformation \cite{Juric2, Juric1, Juric3} is given by 
\begin{equation}\label{twist}
\mathcal F=\text{exp}\left\{ia^\alpha P^\beta\frac{\text{ln}(1+a\cdot P)}{a\cdot P}\otimes M_{\alpha\beta}\right\},
\end{equation}
where $a_{\mu}=\frac{1}{\kappa}u_{\mu}$ is the deformation vector and $u^{\mu}u_{\mu}=0$. This is the so called light-like $\kappa$-deformation of Poincar\'e-Hopf algebra and it is proven in \cite{poster, linreal} that $\mathcal{F}$ satisfies the conditions \eqref{coc} and \eqref{norm}. Now,  the deformed Hopf algebra $(\mathcal{U}^{\mathcal{F}}(\mathfrak{p}), \Delta^{\mathcal{F}}, \epsilon^{\mathcal{F}}, S^{\mathcal{F}}, m^{\mathcal{F}})$  for the generators is given by
\begin{equation}\begin{split}
\Delta^{\mathcal{F}}M_{\mu\nu}&=\mathcal{F}\Delta M_{\mu\nu}\mathcal{F}^{-1}= \Delta M_{\mu\nu}+
(\delta^\alpha_\mu a_\nu-\delta^\alpha_\nu a_\mu)\left(
P^\beta+\frac{1}{2}a^\beta P^2
\right)Z\otimes M_{\alpha\beta},  \\
\Delta^{\mathcal{F}}P_{\mu}&=\mathcal{F}\Delta P_{\mu}\mathcal{F}^{-1}= \Delta P_\mu +\left[
P_\mu a^\alpha - a_\mu
\left( P^\alpha + \frac{1}{2}a^\alpha P^2 \right)Z
\right]\otimes P_\alpha,\\
S^{\mathcal{F}}(M_{\mu\nu})&=\chi S(M_{\mu\nu}) \chi^{-1}= -M_{\mu\nu} +(-a_\mu \delta^\beta_\nu+a_\nu \delta^\beta_\mu) \left(P^\alpha + \frac{1}{2}a^\alpha P^2 \right)M_{\alpha\beta},\\
S^{\mathcal{F}}(P_{\mu})&=\chi S(P_{\mu}) \chi^{-1}=\left[-P_\mu -a_\mu \left(P_\alpha + \frac{1}{2}a_\alpha P^2 \right) P^\alpha \right]Z,\\
\epsilon^{\mathcal{F}}(M_{\mu\nu})&=0, \quad \epsilon^{\mathcal{F}}(P_{\mu})=0,
\end{split}\end{equation}
where $Z=\frac{1}{1+aP}$. This deformed Hopf algebra $(\mathcal{U}^{\mathcal{F}}(\mathfrak{p}), \Delta^{\mathcal{F}}, \epsilon^{\mathcal{F}}, S^{\mathcal{F}}, m^{\mathcal{F}})$ is exactly the $\kappa$-Poincar\'e-Hopf algebra $\mathcal{U}_{\kappa}(\mathfrak{p})$ in the natural realization \cite{Meljanac, Kresic, Meljanac1} (classical basis \cite{Kowalski, Borowiec1}).

Now we consider the algebra $\mathcal{A}_{\star}$ which is induced by the action of the twist \eqref{star}
\begin{equation}\label{star1}
f\star g=\mu\left[\mathcal{F}^{-1}(\triangleright\otimes\triangleright)(f\otimes g)\right], \quad f,g\in\mathcal{A}.
\end{equation}
The algebra of functions on Minkowski space $\mathcal{A}$ is generated by commutative coordinates $x_{\mu}$. Therefore, for the $\star$-product between the coordinates $x_{\mu}$ we have
\begin{equation}\begin{split}\label{xx}
x_{\mu}\star x_{\nu}&=\mu\left[\mathcal{F}^{-1}(\triangleright\otimes\triangleright)(x_{\mu}\otimes x_{\nu})\right] \\
&=x_{\mu}x_{\nu}-ia_{\nu}x_{\mu}-i\eta_{\mu\nu}a^{\alpha}x_{\alpha}.
\end{split}\end{equation}
Using \eqref{xx} we can define a $\star$-commutator between coordinates
\begin{equation}\label{starcom}
[x_{\mu},x_{\nu}]_{\star}\equiv x_{\mu}\star x_{\nu} - x_{\nu}\star x_{\mu}=i(a_{\mu}x_{\nu}-a_{\nu}x_{\mu}).
\end{equation}
Therefore, the noncommutative algebra $\mathcal{A}_{\star}$ is actually the free algebra generated by coordinates $x_{\mu}$ (with the $\star$-product as the multiplication in the algebra)  divided by the two-sided ideal generated by \eqref{starcom}. It is important to note that $\mathcal{A}_{\star}$ is isomorphic\footnote{The isomorphism can be directly obtained by using the realization via Heisenberg algebra (for more details see the Appendix).} as an algebra to the $\kappa$-Minkowski algebra $\hat{\mathcal{A}}$. The $\kappa$-Minkowski algebra $\hat{\mathcal{A}}$ is generated by the noncommuting coordinates $\hat{x}_{\mu}$ that satisfy
\begin{equation}
[\hat{x}_{\mu},\hat{x}_{\nu}]=i(a_{\mu}\hat{x}_{\nu}-a_{\nu}\hat{x}_{\mu}).
\end{equation}

\section{Free scalar field theory in $\kappa$-Minkowski space}

Using the twist \eqref{twist} and the star product \eqref{star1} we want to define an appropriate action functional $\mathcal{S}[\phi]$ for the free scalar field. But before we do that, let us analyze further the properties of the $\star$-product \eqref{star1} and introduce the notion of adjoint. 

The $\star$-product between plane waves is given by
\begin{equation}\label{ee}
\text{e}^{ipx}\star\text{e}^{iqx}=\mu\left[\mathcal{F}^{-1}(\triangleright\otimes\triangleright)(\text{e}^{ipx}\otimes\text{e}^{iqx})\right]=\text{e}^{i\mathcal{D}_{\mu}(p,q)x^{\mu}}
\end{equation}
where 
\begin{equation}\label{D}
\mathcal{D}_{\mu}(p,q)=p_{\mu}(1+aq)+q_{\mu}-a_{\mu}\frac{pq}{1+ap}-\frac{1}{2}a_{\mu}(aq)\frac{p^2}{1+ap},
\end{equation}
and we used the notation $A_{\mu}B^{\mu}=AB$ and $A^{\mu}A_{\mu}=A^2$. The function $\mathcal{D}_{\mu}(p,q)$ defines the deformed momentum addition rule $\mathcal{D}_{\mu}(p,q)=p_{\mu}\oplus q_{\mu}$. It also gives the coproduct for $p_{\mu}$ via
\begin{equation}
\mathcal{D}_{\mu}(p\otimes 1,1\otimes p)=\Delta^{\mathcal{F}}p_{\mu}.
\end{equation} 

In what follows we shall use the results obtained so far in order to define a massive complex scalar field theory on $\kappa$-Minkowski space and to relate it to a corresponding field theory on the undeformed Minkowski space. To construct a proper noncommutative field theory, besides using a star-product, also requires an introduction of the adjoint, i.e. hermitian conjugation operation $\dagger$ (see \cite{Meljanac1} for details). The adjoint of the standard plane wave is defined by
\begin{equation}\label{dag}
(\text{e}^{ipx})^{\dagger}=\text{e}^{iS^{\mathcal{F}}(p)x}.
\end{equation} 
We assume that the standard commutative complex scalar field has a Fourier expansion
\begin{equation}\label{four}
\phi(x)=\int \d^{n}p\ \tilde{\phi}(p)\text{e}^{ipx}.
\end{equation}
Now we can write the action for a non-interacting massive complex scalar field as 
\begin{equation}\begin{split}\label{action}
\mathcal{S}[\phi]&=\int \d^n x\ \mathcal{L}(\phi, \partial_{\mu}\phi)\\
&=\int \d^n x\ (\partial_{\mu}\phi)^{\dagger}\star (\partial^{\mu}\phi) +m^2\int \d^n x\ \phi^{\dagger}\star\phi .
\end{split}\end{equation}
To find the relation with the corresponding commutative action on undeformed Minkowski space we shall first analyze 
\begin{equation}
\int \d^n x\ \varphi^{\dagger}\star\phi
\end{equation}
 for all $\varphi, \phi\in\mathcal{A}$. According to \eqref{dag} and \eqref{four} the Fourier expansion of $\varphi^{\dagger}$ is
\begin{equation}\label{four1}
\varphi^{\dagger}(x)=\int \d^{n}p\ \tilde{\varphi}^*(p)\text{e}^{iS^{\mathcal{F}}(p)x},
\end{equation}
where $*$ denotes the standard complex conjugation. From \eqref{ee}, \eqref{D}, \eqref{four} and \eqref{four1} we have
\begin{equation}\begin{split}\label{racun}
\int \d^{n}x\ \varphi^{\dagger}\star\phi&=\int \d^{n}x\int \d^{n}p\int \d^{n}q\ \tilde{\varphi}^*(p) \tilde{\phi}(q)\text{e}^{iS^{\mathcal{F}}(p)x}\star\text{e}^{iqx}\\
&=\int \d^{n}p\int \d^{n}q\ \tilde{\varphi}^*(p) \tilde{\phi}(q)\int \d^{n}x\ \text{e}^{i\mathcal{D}_{\mu}\left(S^{\mathcal{F}}(p),q\right)x^{\mu}}\\
&=(2\pi)^n\int \d^{n}p\int \d^{n}q\ \tilde{\varphi}^*(p) \tilde{\phi}(q)\  \delta^{(n)}\left[\mathcal{D}\left(S^{\mathcal{F}}(p),q\right)\right].
\end{split}\end{equation}
To calculate the $\delta^{(n)}$-function we use the identity
\begin{equation}
\delta^{(n)}\left(F(p,q)\right)=\sum_{i}\frac{\delta^{(n)}(q-q_i)}{\left|\det\left(\frac{\partial F_{\mu}(p,q)}{\partial q_\nu}\right)\right|_{q=q_i}}.
\end{equation}
It can be shown  that \cite{Meljanac1}
\begin{equation}
\delta^{(n)}\left(\mathcal{D}(S^{\mathcal{F}}(p),q)\right)=\sqrt{1+a^2p^2}\ \delta^{(n)}(p-q),
\end{equation}
where
\begin{equation}
\delta^{(n)}(p-q)=\frac{1}{(2\pi)^n}\int \d^n x\ \text{e}^{i(p-q)x}.
\end{equation}
However, since in our case we are dealing only with the light-like deformations, i.e. $a^2=0$, we have that 
\begin{equation}
\int \d^{n}x\ \text{e}^{iS^{\mathcal{F}}(p)x}\star\text{e}^{iqx} =(2\pi)^n\delta^{(n)}(p-q).
\end{equation}
Thus, finally\footnote{For the general deformation $a_{\mu}$ we would have $\int \d^{n}x\ \varphi^{\dagger}\star\phi=\int \d^{n}x\ \varphi^{*}\sqrt{1-a^2\partial^2}\phi$ (for more details see \cite{Meljanac1}).}
\begin{equation}\label{rez}
\int \d^{n}x\ \varphi^{\dagger}\star\phi=\int \d^{n}x\ \varphi^{*}\phi.
\end{equation}
We see that under the integration sign we can simply drop out the $\star$-product and replace it with an ordinary pointwise multiplication. This means that the free massive scalar field theory on $\kappa$-Minkowski space loses a nonlocal character and takes on an undeformed shape (at least for the mass term).

Now we have to investigate the partial integration for $\kappa$-Minkowski space using $\star$-product. By going through the same steps as in \eqref{racun}, that brought us to the result \eqref{rez}, we get the requred partial integration formula\footnote{In the case that the derivative is not affected by the adjoint action we would have $$\int \d^{n}x\ \partial_{\mu}\varphi^{\dagger}\star\phi=\int \d^{n}x\ \varphi^{\dagger}\star S^{\mathcal{F}}(\partial_{\mu})\phi=\int \d^{n}x\ \varphi^{*}S^{\mathcal{F}}(\partial_{\mu})\phi,$$ which then leads to a deformed Klein-Gordon operator that is not the Casimir operator of $\kappa$-Poincar\'e algebra \cite{Meljanac1}.}
\begin{equation}\label{rez1}
\int \d^{n}x\ (\partial_{\mu}\varphi)^{\dagger}\star\phi=-\int \d^{n}x\ \varphi^{\dagger}\star\partial_{\mu}\phi=-\int \d^{n}x\ \varphi^{*}\partial_{\mu}\phi.
\end{equation}
Using the properties of the $\star$-product \eqref{rez} and \eqref{rez1} and the expression for the action \eqref{action}, we have
\begin{equation}\begin{split}\label{S}
\mathcal{S}[\phi]&=\int \d^n x\ (\partial_{\mu}\phi)^{\dagger}\star (\partial^{\mu}\phi) +m^2\int \d^n x\ \phi^{\dagger}\star\phi \\
&=-\int \d^{n}x\ \phi^{*}\partial_{\mu}\partial^{\mu}\phi+m^2\int \d^{n}x\ \phi^{*}\phi\\
&=\int \d^{n}x\left[(\partial_{\mu}\phi)^* (\partial^{\mu}\phi)+m^2\phi^* \phi\right].
\end{split}\end{equation}
Eq.~\eqref{S} illustrates the equivalence of the noncommutative free field theory with the commutative one. Action \eqref{S} is Poincar\'e invariant and it should be noted that the Klein-Gordon operator is undeformed, i.e. $-\partial_{\mu}\partial^{\mu}+m^2$.

\section{The $\phi^4$ scalar model on $\kappa$-Minkowski space}

In the previous section we have investigated the free scalar field in the $\kappa$-Minkowski space and now we want to introduce the interaction between them. In order to introduce the self-interacting $\phi^4$ terms into the action functional, we should replace the pointwise multiplication with the $\star$-product. This means that we should incorporate six terms corresponding to all possible permutations of fields $\phi$ and $\phi^\dagger$, but due to the properties of the $\star$-product under the integration sign, \eqref{rez}, these six permutations can be reduced to just two mutually nonequivalent terms and we can write the action for the noncommutative $\phi^4$ model as 
\begin{equation}\begin{split}\label{S0}
\mathcal{S}[\phi]&=\int \d^n x\ (\partial_{\mu}\phi)^{\dagger}\star (\partial^{\mu}\phi) +m^2\int \d^n x\ \phi^{\dagger}\star\phi \\
&+\frac{\lambda}{8}\int\d^{n} x\left(\phi^\dagger \star \phi^\dagger \star \phi \star \phi+\phi^{\dagger}\star\phi\star\phi^{\dagger}\star\phi\right).
\end{split}\end{equation}
The first part of the action reduces to the free commutative situation \eqref{S}, while the interaction part of the action is very complicated and here we  use \eqref{star1}, \eqref{ee}, \eqref{D}, \eqref{rez} and \eqref{rez1}
in order to expand the interaction part in the deformation parameter $\frac{1}{\kappa}$ and to further simplify the expression for the action functional. Finally we get
\begin{equation}\begin{split}\label{S1}
\mathcal{S}[\phi]=\int \d^{n}x&\left[(\partial_{\mu}\phi)^* (\partial^{\mu}\phi)+m^2\phi^* \phi+\frac{\lambda}{4}(\phi^* \phi)^2\right]\\
+i\frac{\lambda}{4}\int \d^{n}x&\bigg[a_{\mu}x^{\mu}\left((\phi^{*})^2(\partial_{\nu}\phi)\partial^{\nu}\phi-\phi^2(\partial_{\nu}\phi^*)\partial^{\nu}\phi^*\right)\\
&+a_{\nu}x^{\mu}\left(\phi^2(\partial_{\mu}\phi^*)\partial^{\nu}\phi^*-(\phi^*)^2 (\partial_{\mu}\phi)\partial^{\nu}\phi\right)\\
&+\frac{1}{2}a_{\nu}x^{\mu}\phi^* \phi\left((\partial_{\mu}\phi^\dagger)\partial^{\nu}\phi-(\partial_{\mu})\partial^{\nu}\phi^*\right)\bigg] +\mathcal{O}\left(\frac{1}{\kappa^2}\right).
\end{split}\end{equation}
The action obtained in \eqref{S1} via twist $\mathcal{F}$ \eqref{twist} is exactly the same as in \cite{trampetic} where the theory of realizations was used \cite{Kresic, Meljanac, Meljanac1}. 

In \cite{trampetic} it is argued that the changes in the particle statistics (due to twist $\mathcal{F}$) are encoded in the nonabelian momentum addition law \cite{Freidel, Kowalski1, Kowalski2, Lukierski1}. It can be further shown that the accordingly induced deformation of $\delta$-function yields the usual $\delta$-function multiplied by a certain statistical factor. Thus, in order to obtain Feynman rules in the momentum space one can use the following two  steps:
\begin{enumerate}
\item Use the standard commutative methods of QFT in obtaining the propagators and vertex by treating the noncommutative contribution to the action \eqref{S1} just as a small perturbation.
\item Include the twisted nature of particle statistics. Namely, the statistics of particles is twisted and this can be implemented by requiring that the ordinary addition/subtraction rules for the particle momentum are deformed. Particularly, this means that we have 
\begin{equation}\label{add}
\sum_{i}k^{\mu}_{i}\longrightarrow\sum_{\oplus i}k^{\mu}_{i}\ \ \&\ \ \sum_{i}k^{\mu}_{i}-\sum_{j}p^{\mu}_{j}\longrightarrow\sum_{\oplus i}k^{\mu}_{i}\ominus\sum_{\oplus j}p^{\mu}_{j}
\end{equation}
where $p_{\mu}\oplus q_{\mu}\equiv\mathcal{D}_{\mu}(p,q)$ and $p_{\mu}\ominus q_{\mu}\equiv p_{\mu}\oplus S^{\mathcal{F}}_{\mu}(q)$. Here ${D}_{\mu}(p,q)$ is given by \eqref{D} and note that due to the relation that exists between the coproduct and antipode we have $p_{\mu}\oplus S^{\mathcal{F}}_{\mu}(p)=0$ (this is induced by the underlining Hopf algebra structure).
\end{enumerate}

Using the above explained idea, the two-point connected Green's function up to one loop corrections takes the following form
\begin{equation}\label{G}
G(k_1,k_2)=(2\pi)^n \delta^{(n)}(k_1-k_2)\frac{i}{k^2_1+m^2-\Pi(k_1)},
\end{equation}
where
\begin{equation}\label{pi}
\Pi(k_1)=\frac{\lambda m^2}{32\pi^2}\left[(1+3ak_1)\left(\frac{2}{\epsilon}+\psi(2)+\ln\frac{4\pi\mu^2}{m^2}\right)-\frac{9}{4}ak_1\right]
\end{equation}
is the tadpole diagram contribution to the self-energy of the scalar particle \cite{trampetic}.
Here we used the Euclidean prescription and the $n=4-\epsilon$ dimensional regularization\footnote{For more details see \cite{trampetic}.}. After introducing the appropriate counterterm, we get the finite two-point function
\begin{equation}
G_{fin}(k_1,k_2)=\frac{G}{1+G\frac{\lambda m^2}{32\pi^2}\left[f-(1+3ak_1)\left(\frac{2}{\epsilon}+\psi(2)+\ln\frac{4\pi\mu^2}{m^2}-\frac{9}{4}\frac{ak_1}{1+3ak_1}\right)\right]},
\end{equation}
where
\begin{equation}
G=\delta^{(n)}(k_1 -k_2)\frac{i}{k^2_1+m^2}
\end{equation} 
and $f$ is an arbitrary dimensionless function fixed by the renormalization condition.

\section{Final remarks}
The integral measure problems on $\kappa$-Minkowski space are avoided by using the Drinfeld twist operator  \eqref{twist} for the light-like $\kappa$-deformation of Poincar\'e-Hopf algebra. We have shown that the $\star$-product obtained from the twist \eqref{twist} has very nice properties under the integration sign and enables us to define a free scalar field theory on $\kappa$-Minkowski space that is equivalent to a commutative one on a usual Minkowski space. 

We present the first interacting scalar field theory compatible with $\kappa$-Poincar\'e-Hopf algebra obtained via Drinfeld twist \eqref{twist}. The action functional \eqref{S0} defines a $\phi^4$ scalar model on $\kappa$-Minkowski space. The truncated $\kappa$-deformed action \eqref{S1} does not have the UV/IR mixing \cite{Grosse}.

In the computation of the two-point function we have implemented the \textsl{hybrid} approach \cite{trampetic}, i.e. we have used the deformed momentum addition/subtraction \eqref{add}.  There are new non-vanishing contributions even in $n=4$ dimensions. They are arising from the $\kappa$-deformed momentum addition/subtraction and the corresponding energy-momentum conservation rule stemming from the deformed $\delta$-function \cite{trampetic}. For the conserved external momenta, i.e. $k_1=k_2$ we obtained the two-point function \eqref{G}, where the finite parts represent the modification of the scalar field self-energy $\Pi$ \eqref{pi} and depend explicitly on the regularization parameter $\mu^2$ and the mass of the scalar field $m^2$. It is important to note that \eqref{pi} contains the finite correction $ak_1$ due to the deformed statistics on the $\kappa$-Minkowski space, thus, we obtained the explicit dependance on the direction of the propagation energy, its scale $\left|k_1\right|=E$ and the deformation parameter $\frac{1}{\kappa}$.

In this paper we have used the truncated action \eqref{S1}, but if we could  find the $\star$-product \eqref{star1} in closed form, we could express the interacting part of the action \eqref{S0} up to all orders in $\frac{1}{\kappa}$ and try to obtain a one loop correction to the two-point function valid in all orders in $\frac{1}{\kappa}$. 
So far we have analyzed the \textsl{hybrid} version of the $\phi^4$ scalar field theory. The next step would be to investigate  the interacting  quantum field theory on $\kappa$-Minkowski space, but with the proper notion of twisted particle statistics. In order to do this , we have to investigate the $R$-matrix \cite{jhep} which would modify the quantization procedure, that is we have to modify the algebra of creation and annihilation operators via
\begin{equation}
\phi(x)\otimes\phi(y)-R\phi(y)\otimes\phi(x)=0
\end{equation}
which would modify the usual spin-statistics relations of the free bosons at Planck scale. The $R$-matrix is defined by the twist operator $R=\tilde{\mathcal{F}}\mathcal{F}^{-1}$. The idea is that the $R$-matrix would enable us to define particle statistics and to properly quantize fields, while the twist operator would provide the star-product, which is crucial for writing the action in terms of commutative variables. In the end, this would enable us to derive the Feynman rules and find the noncommutative correction to the propagator and the vertex to all orders in $\frac{1}{\kappa}$. 

In order to investigate gauge field theories on $\kappa$-Minkowski space, the notion of noncommutative differential geometry is necessary. The classification of all bicovariant differential calculi on $\kappa$-Minkowski space and the concept of classical noncommutative field theory was developed in \cite{Juric3}. This was obtained by embedding the algebra of coordinates and one-forms into a super-Heisenberg algebra. It will be very interesting to apply the Drinfeld twist \eqref{twist} for the light-like $\kappa$-Poincar\'e algebra to obtain all aspects of noncommutative differential geometry and to formulate a noncommutative gauge field theory.

\ack
The work by T.~J. and S.~M.  has been fully supported by Croatian Science Foundation under the project (IP-2014-09-9582).
The work by A.S. was supported by the European Commission and the Croatian Ministry of Science, Education and Sports through grant project financed under the Marie Curie FP7-PEOPLE-2011-COFUND, project NEWFELPRO.

\appendix
\section{Realization via Heisenberg algebra}
The realization of the $\kappa$-Minkowski space is defined by the unique map from the algebra $\mathcal{A}$ of functions $f$ in commuting coordinates $x_{\mu}$ to the algebra $\hat{\mathcal{A}}$ of noncommuting ``functions'' $\hat{f}$ generated by noncommutative coordinates $\hat{x}_{\mu}$. This map $\Omega:\mathcal{A}\longrightarrow\hat{\mathcal{A}}$ is uniquely characterized by $f\mapsto\hat{f}$ such that
\begin{equation}
\hat{f}\triangleright 1=f.
\end{equation}
The action $\triangleright:\mathcal{H}\otimes\mathcal{A}\longrightarrow\mathcal{A}$ is defined by 
\begin{equation}
p_{\mu}\triangleright 1=0, \quad x_{\mu}\triangleright 1=x_{\mu}, \quad p_{\mu}\triangleright x_{\nu}=-i\eta_{\mu\nu}1,
\end{equation}
and $\mathcal{H}$ is the Heisenberg algebra generated by $x_{\mu}$ and $p_{\mu}$ such that
\begin{equation}
[x_{\mu},x_{\nu}]=[p_{\mu},p_{\nu}]=0, \quad [p_{\mu}, x_{\nu}]=-i\eta_{\mu\nu}1.
\end{equation}
We can find the explicit realization of noncommuting coordinates $\hat{x}_{\mu}$ and Poincar\'e generators $M_{\mu\nu}$ and $P_{\mu}$ via generators of $\mathcal{H}$
\begin{equation}\begin{split}
\hat{x}_{\mu}&=x_{\mu}+a^{\alpha}M_{\mu\alpha},\\
M_{\mu\nu}&=x_{\mu}p_{\nu}-x_{\nu}p_{\mu},\\
P_{\mu}&=p_{\mu}.
\end{split}\end{equation}
The star product is given by
\begin{equation}
f\star g=\hat{f}\hat{g}\triangleright 1=\hat{f}\triangleright \hat{g},
\end{equation}
where $\hat{f}\triangleright 1=f$ and $\hat{g}\triangleright 1=g$. This gives the isomorphism $\hat{\mathcal{A}}\cong\mathcal{A}_{\star}$.\\

\end{document}